\documentclass[manuscript]{emulateapj}
\usepackage{graphicx}
\usepackage{amsmath}
\usepackage{amssymb}
\usepackage{wasysym}
\usepackage{color}

\shorttitle{LAMOST BHBs I}
\shortauthors{Vickers et al.}

\begin{document}

\author{John J. Vickers\altaffilmark{$\dagger$, 1, 2}, Zhao-Yu Li\altaffilmark{$\dagger$, 1, 2}, Martin C. Smith\altaffilmark{3}, Juntai Shen\altaffilmark{$\dagger$, 1, 2, 3}}

\title{A LAMOST BHB Catalog and Kinematics Therein I: Catalog and Halo Properties}

\altaffiltext{1}{Department of Astronomy, School of Physics and Astronomy, Shanghai Jiao Tong University, 800 Dongchuan Road, Shanghai 200240, People's Republic of China}

\altaffiltext{2}{Key Laboratory for Particle Astrophysics and Cosmology (MOE) / Shanghai Key Laboratory for Particle Physics and Cosmology, Shanghai 200240, China}

\altaffiltext{3}{Key Laboratory for Research in Galaxies and Cosmology, Shanghai Astronomical Observatory, Chinese Academy of Sciences, 80 Nandan Road, Shanghai 200030, People's Republic of China}

\altaffiltext{$\dagger$}{\\ johnjvickers@sjtu.edu.cn \\ lizy.astro@sjtu.edu.cn \\ jtshen@sjtu.edu.cn}

\def\mean#1{\left< #1 \right>}

\renewcommand{\thempfootnote}{\arabic{mpfootnote}}

\begin{abstract}

  We collect a sample of stars observed both in LAMOST and Gaia which have colors implying a temperature hotter than 7000 K. We train a machine learning algorithm on LAMOST spectroscopic data which has been tagged with stellar classifications and metallicities, and use this machine to construct a catalog of Blue Horizontal Branch stars (BHBs) with metallicity information. Another machine is trained using Gaia parallaxes to predict absolute magnitudes for these stars. The final catalog of 13,693 BHBs is thought to be about 86\% pure, with $\sigma_{[Fe/H]}\sim$0.35 dex and $\sigma_{G}\sim$0.31 mag. These values are confirmed via comparison to globular clusters, although a covariance error seems to affect our magnitude and abundance estimates.

  We analyze a subset of this catalog in the Galactic Halo. We find that BHB populations in the outer halo appear redder, which could imply a younger population, and that the metallicity gradient is relatively flat around [Fe/H] = -1.9 dex over our sample footprint. We find that our metal rich BHB stars are on more radial velocity dispersion dominated orbits ($\beta \sim 0.70$) at all radii than our metal poor BHB stars ($\beta \sim 0.62$).

\end{abstract}

\section{Introduction}\label{sec:introduction}

Blue horizontal branch stars are reliable ``standard candle'' stars commonly used for studying the Galactic halo (e.g. \citealt{gre1974}, \citealt{bee1992}, \citealt{yan2000}). They are helium burning giants, which have evolved off the red giant branch (and so, thought to be old, \citealt{hoy1955}), with absolute magnitudes near 0 in a variety of optical bands. The anatomy of the horizontal branch begins at the cool end with the red horizontal branch (near the red clump), progresses blueward through the RR Lyrae gap, into the blue horizontal branch, and then falls off in magnitude down the extreme horizontal branch (see the review of \citealt{cat2009}).

Owing to their low surface gravities, BHBs exhibit narrow spectral lines compared to main sequence stars and may be selected on that basis (\citealt{pie1983}, \citealt{fly1994}, \citealt{cle2002}). This variation in spectral line shape, near the absorption line series' limits, also causes a net change in continuum flux, which may be exploited to select them based on filter colors (for example: \citealt{bel2010}, \citealt{dea2011}, \citealt{vic2012}).

While photometric studies of BHB stars allow observations extending to great distances (the Magellanic Clouds, \citealt{bel2016}, or the halo to hundreds of kpc, \citealt{dea2018ii}, \citealt{fuk2018}, \citealt{fuk2019}, \citealt{nie2015}, \citealt{tho2018} for example), they suffer worse contamination ($\sim 30\%$; \citealt{sir2004}, \citealt{bel2010}) than spectroscopic studies ($<$10\%; \citealt{xue2008}, \citealt{ruh2011}; although, note that impressive progress has been made with specialized photometric filters, see for example \citealt{sta2019} who find their photometrically-selected sample of BHBs to be both more than 90\% pure, and more than 90\% complete)\footnote{``Purity'' also called ``precision'' refers to the number of true positives out of all positive classifications. ``Contamination'' refers to the number false positive misclassifications out of all positive classifications (one minus the purity).  ``Completeness,'' also called ``recall,'' refers to the number of ``true positive'' classifications out of all true positives (that is, the proportion of true candidates which are successfully recovered).}. Spectroscopy also unlocks the 6-th phase coordinate through radial velocity measurements.

Unfortunately, BHB stars' hot temperatures ($>$ 7000 K) necessitate special spectroscopic treatment (for an example, see the Sloan Digital Sky Survey pipeline, \citealt{lee2008}, and the special A-star procedure they implemented, \citealt{wil1999}), compared to the standard pipeline, which is generally optimized for FGK dwarfs (for example: \citealt{boe2018}, \citealt{lee2015}), so specialized methods must be coded.

In recent years, however, spectroscopic analysis has benefited disproportionately from machine learning techniques. Learning algorithms may use examples of data which have already been ``labeled'' with spectroscopic parameters to label new data, without the need for researchers to code specific procedures for different species of stars by hand (e.g. ``The Cannon'' \citealt{nes2015} and the ``Stellar LAbel Machine'' SLAM \citealt{zha2020}). This approach allows for many different types of stars to be reduced using a single, generalized, pipeline procedure, with the machine learning how to deal with various types of stars from previously reduced data.

Using such an approach we will construct one of the largest spectroscopic samples of BHB stars to date, almost 14,000 BHB candidates, which will be used to investigate the ancient history of the Milky Way.

BHBs have long been used to study the Milky Way halo. Their bright absolute magnitudes allow them to probe great distances, and their old ages are ideal for studying the canonical in-situ halo. While the halo formation scenario has been more complicated in recent years from the monolithic collapse of \citet{egg1962}, to the present understanding of heirarchical formation (\citealt{sea1978}, \citealt{hel1999}, \citealt{new2002}), BHBs remain useful as the accreted satellites often possess ancient BHB populations of their own.

This paper proceeds as follows: in Section \ref{sec:xgb} we describe our data to be classified and reduced, and go through how we assign classifications, metallicities, and absolute magnitudes to our data; in Section \ref{sec:verify} we describe some ``sanity checks'' to double check our reduced data (such as comparing pipeline values to globular cluster values); in Section \ref{sec:coords} we describe our coordinate systems; in Section \ref{sec:analysis} we look over the chemistry, kinematics, and ages of our halo BHBs; we discuss our findings in Section \ref{sec:discuss} and conclude in Section \ref{sec:conclusion}.

\subsection{Abbreviations and Notation}
For this paper, we will use the following abbreviations: ``BHB'' is a Blue Horizontal Branch star; ``MSA'' is a Main Sequence A-type star\footnote{For this paper, a blue straggler star will be considered a MSA star as they are similar in observational qualities.}; ``XGB'' indicates an eXtreme Gradient Boosted (XGBoost) random forest algorithm or a value which has been predicted by one; $\varpi$ indicates Gaia parallax or a value calculated from it; the subscript ``0'' indicates a photometric value which has been corrected using the full column extinction of \citet{sch1998} with the filter coefficients from \citet{cas2018}; an upper-case G will represent an absolute magnitude in the Gaia g band; ``SNR'' stands for signal-to-noise ratio.

\section{Predicting Stellar Types, Metallicities, and Absolute Magnitudes with XGBoost}\label{sec:xgb}

\subsection{Data}\label{ssec:data}

Our BHB catalog will be constructed from data collected by the Large Area Multi Object Spectroscopic Telescope (LAMOST; \citealt{luo2015}). This telescope, located at Xinglong Station in Hebei Province, collects 4000 spectra per exposure at a resolution of R$\sim$1800 down to a magnitude of r=19. As of data release 5, the survey contains about 9 million spectra in total.

The standard LAMOST pipeline \citep{wu2011} struggles to give accurate stellar parameters for stars hotter than $\sim$7000 K, as it is optimized for the more prevalent main sequence stars. Since we are interested in BHB stars, which have temperatures up to 10,000 K, we will need to process the spectra ourselves. 

155,549 LAMOST spectra are selected as BHB candidates to have Gaia DR2 (\citealt{lin1996}, \citealt{gai2018}) colors bluer than (bp-rp) = 0.5, roughly corresponding to 7000 K. These spectra are normalized following \citet{boe2018}.

\subsection{Foundation}\label{ssec:foundation}
The fundamental spectral difference between BHB stars and the main contaminants in this spectral color regime (main sequence A stars and blue stragglers, quasars, and white dwarfs) are the presence of hydrogen absorption lines (which are instead emission lines in quasar spectra) which are narrow due to the low surface gravity of giant BHBs compared to the other stellar contaminants. See, for example, \citet{vic2012} Figures 7 and 8, \citet{yan2000} Figure 8.

Previous works frequently relied on fitting line profiles to discriminate BHBs from different species of stars by their surface gravity (e.g. \citealt{cle2002}, \citealt{sir2004}, \citealt{xue2008}). However, with modern machine learning tools becoming readily available and easily accessible on personal computers, we opt to classify our BHB stars using a data-driven algorithm.

The benefit of using a machine learning framework to classify BHB stars lies in the fact that we can readily use information from the entire spectra, whereas prior line-fitting methods rely on carefully selecting the wavelength ranges of individual lines. The absorption features follow profiles\footnote{Sometimes the profile is fit with a Voigt profile, which is a combination of a Gaussian core and Lorentz wings, representative of the differing effects on spectral line shape of abundance and gravity.} which merge into the continuum, or even into each other as is the case in stars with exceptionally strong surface gravities, such as white dwarfs. Utilizing the entire spectra avoids problems associated with mis-selecting profile extents and shape profiles, and allows more spectral information, such as non-targetted, incidental lines, to be utilized.

The difficulty of using a  machine learning supervised algorithm is that we require a training set which is regarded as ``true" for the machine to learn from. The supervised algorithm will consider a training data set which is in the same data format as the data to be classified, (i.e. LAMOST spectra) and which has tags for the classes (i.e. ``BHB,'' ``not-BHB''), and construct a mathematical equation\footnote{In most cases, some techniques such as those based on decision trees construct a series of yes-no questions.} through which the data may be passed to yield a numerical value. This numerical value may be a regressed value (as is the case when fitting a continuous metric, such as metallicity), or a 1-0 class value (as we are doing now for BHB, not-BHB).

This need for a training set means that a subset of the data must be processed and classified by hand before the bulk of the data may be classified by machine -- it must be ``supervised" by an intelligent actor. In this way, modern data-driven pipelines rest very directly on the shoulders of past researchers who have painstakingly reduced subsets of these data by hand.

\subsection{Training Data}\label{ssec:training_data}

The Century Survey of \citet{bro2008} is a catalog of 2414 color-selected BHB candidates over 10\% of the sky from Two-Micron All Sky Survey \citep{skr2006} in 12.5 $\leq$ $J_{0}$ $\leq$ 15.5. This survey confirmed 655 of these candidates as ``true'' BHB stars. LAMOST has observed 573 of the 2414 candidates, and 213 of the 655 confirmed BHB stars (at SNR$_{g, r, i}$ $>$ 10).

The catalog of \citet{xue2008} took Sloan Digital Sky Survey \citep{yor2000} spectra and reprocessed $\sim$10,000 color selected BHB candidates using a specialized pipeline for hot stars. This yielded 2558 confirmed BHB stars. LAMOST has observed 833 of these 10,000 candidates, and 380 of the 2558 confirmed BHB stars (at SNR$_{g, r, i}$ $>$ 10).

From these two surveys, we have a training set of 1,406 objects in LAMOST: 593 BHBs and 813 not-BHBs. To prepare the data to be interpreted, we shift all the spectra to their rest wavelengths according to their LAMOST radial velocity. Since the spectra are non-linearly sampled in observed wavelength (R = $\lambda$/$\Delta\lambda$ $\sim$ 1800), the rest wavelength spectra are sampled incongruently with each other. To be directly compared, the data must be sampled at the same rest wavelengths. This may be accomplished by binning the data, which loses information but conserves computer resources, or by upsampling the data (linearly interpolating between the observed wavelengths at set points on a wavelength grid at a higher frequency than the original data), which conserves information but becomes more computationally costly. We opt for the latter, with an upsampled grid every 0.84 $\AA$, which is slightly more than one grid-point per measurement at the blue end of the spectra and more dense per measurement at the red end. Our grid extends from 3850 to 8900 $\AA$, which is the maximal extent covered by all spectra in our data set (at rest wavelengths). Note that this data set is pre-cut to include only data with radial velocities less than 1000 km s$^{-1}$.\footnote{Large redshifts artificially compress our rest spectra to be smaller than the grid we use. This cut probably removes spectra of blue items at large distances, such as quasars.}

Our training sample, using data from \citet{xue2008} and the Century Survey \citep{bro2008}, contains 1,406 stars with BHBs making up 42\%. For a full breakdown, see Table \ref{tab:training_set}.

\begin{table*}
  \caption{}
  \begin{center}
  \begin{tabular}{ | c | c | c | c | }
    \hline
    Survey (Total / $<$10\%$\varpi$ error) & & N$_{Full}$ & N$_{10\%\varpi}$ \\
    \hline
    Century (573/175) & BHB & 213 (37\%) & 18 (10\%) \\
    & MSA & 154 (27\%) & 68 (39\%) \\
    & BHB/MSA & 47 (8\%) & 4 (2\%) \\
    & subdwarf & 17 (3\%) & 12 (7\%) \\
    & DA white dwarf & 8 (1\%) & 8 (5\%) \\
    & None & 134 (23\%) & 65 (37\%) \\
    \hline
    Xue (833/180) & BHB & 380 (46\%) & 103 (57\%) \\
    & MSA/BS & 453 (54\%) & 77 (43\%) \\
    \hline
  \end{tabular}
\tablecomments{
The makeup of our training sets in the full sample and in the 10\% parallax error sample.
}
\label{tab:training_set}
\end{center}
\end{table*}

\subsection{The Machine}\label{ssec:xgb}
With our training data and data to be classified in hand, we return to the machine learning problem. We utilize the eXtreme Gradient Boosting method (XGB\footnote{https://xgboost.readthedocs.io/}, \citealt{che2016}). This modern machine learning method is an optimized version of ``gradient boosting'' \citep{fri2001}, which was found to be a generalized version of the popular ``adaptive boosting'' \citep{fre2009} method, which belongs to the ``random forest'' \citep{ho1995} family of algorithms. A random forest is a series of weak decision trees. These trees each calculate a value Y based on input characteristics X for a data set by asking a series of yes or no questions; they are kept weak by only allowing each tree to ask a limited number of questions and only showing each tree a random subset of the data instead of the whole set\footnote{A problem common to decision trees, and a risk inherent in all supervised machine learning techniques, is so-called ``overtraining,'' whereby the machine memorizes the training data exactly. This provides excellent precision and recall of the training data, but poor metrics for unseen testing data. This is why the individual components must be kept ``weak,'' however the resulting conglomerate will be ``strong.''}. The trees then vote as a forest on a value of Y for each data array X. An ``adaptive boosted'' forest is planted one tree at a time instead of randomly all at once. Each new tree is chosen to correct classifications or regressions that the existing forest has failed at using weighting, so the forest tends to grow more accurate with each new tree. ``Gradient boosting'' is a more generalized method of this selective tree planting method, and ``eXtreme Gradient Boosting'' is a computer-optimized implementation of this generalized method.

The machine requires an array of X values (the spectroscopic fluxes, shifted and upsampled as described above) and a list of classes or values (for example: Y${Class}$ $=$ BHB or not-BHB, or Y$_{[Fe/H]}$ = -1.2 or -0.7).

We create three machines: the first machine classifies stars as BHB or not-BHB, using the training data (1,406 stars, 593 of which are BHBs); the second machine regresses metallicities using the same catalogs, but uses only BHB estimated metallicities (that is, a training set with 552 BHBs, extending from [Fe/H] of -3.0 to -0.3, the 41 BHB difference is from those which do not have metallicities assigned in our training data\footnote{The 41 BHBs with missing metallicity values all occur in the catalog of \citet{xue2008}. In their work, they classified BHB stars based on a line-shape method which they implemented themselves on the SDSS spectra, and the reported metallicities were collected from a separate calculation, the WBG pipeline in SDSS. The WBG, or ``Wilhelm, Beers, and Gray'' pipeline, is optimized for the treatment of hot stars, such as BHBs, and the details may be found in \citet{wil1999}. It is unclear why the SDSS pipeline was unable to assign a metallicity to these stars while \citet{xue2008} were able to calculate their line parameters for classification purposes. We note that the stars with missing values seem to be preferentially bluer and dimmer than the bulk of the \citet{xue2008} sample.}); the third machine regresses absolute magnitudes using the parallax $\varpi$ for a training set of 697 BHBs (extending from G of -1.2 to 2.5, this training set is constructed from LAMOST-observed stars identified as BHBs from the prior classification machine, which also have good parallax measurements, and is unrelated to the training sets constructed from the Century Survey or \citealt{xue2008} sample).

\subsubsection{Note on Predicting Absolute Magnitudes}\label{sssec:disk_height}
One of the desirable characteristics of BHB stars is their bright and stable absolute magnitudes. These absolute magnitudes are sometimes treated as a constant, and sometimes treated as a function of color (see, for example, \citealt{dea2011}). We use the XGB algorithm to predict a value, which should offer more flexibility than a flat value or a color-magnitude fit.

Absolute magnitudes are not given in our training catalogs. To predict absolute magnitudes, we will need a training set of BHB stars with ``known'' absolute magnitudes. We can easily calculate this for BHB stars which have good parallax measurements and with correct apparent magnitudes. To ensure correct parallaxes, we use stars with 10\% parallax errors or less. To ensure we have the correct apparent magnitudes, we correct our data for extinction using the maps of \citet{sch1998} and only use data which is 750 pc or more outside of the plane. This should ensure that the bulk of the dust column from the reddening map is between us and the star and we are not over-correcting our data. This training set of 697 BHBs is used to predict the absolute magnitudes of the rest of the BHBs in our sample.

\subsection{Note on the Parallax Offset}

Note that we have not used a parallax offset (for an inclusive overview of the field, see \citealt{zin2019} and references therein). The reasoning for this is that a parallax offset of 0.054 mas (a reasonably central value from \citealt{sch2019} when compared to various studies shown in Figure 1 of \citealt{zin2019}), for example, shifts the XGB estimated absolute magnitudes of our BHBs by approximately 0.3 magnitudes. Consider for illustration, a BHB observed with a magnitude of 12 and a parallax of 0.4 mas, which would have an absolute magnitude of 0.01; increasing the parallax to 0.45 mas would shift the absolute magnitude by 0.25 to $G$ = 0.26. This shift causes our distance comparison to globular clusters, as seen in Section \ref{sec:globs} and Figure \ref{fig:glob_bhb}, to be systemically offset with distance moduli too small (distances closer than their globular cluster's).

We suggest that our sample of objects used to train our data (having colors $b$-$r$ $<$ 0.5) may have a different, or negligible, Gaia parallax offset. See for example the discussion in \citet{zin2019} and their Figures 6 and 7, which shows a parallax offset trendline in color and temperature and which may approach zero when considering such hot objects (although the trendline is complex and poorly behaved, and their data do not extend to this color regime).

We have attempted an iterative search of the Gaia parallax offset values which would produce estimated distance moduli closest to the distance moduli of our globular cluster sample, but the improvement over a zero-offset was minor. We feel that such a topic deserves a more thorough investigation than we have given it, but such an investigation would distract from the main thrust of this current work. We simply say that, for our specific sample, it seems unnecessary and perhaps counterproductive to include the Gaia parallax zero-point offset. We do, however, encourage anyone using the absolute magnitudes presented in our catalog to investigate this for themselves. The resulting magnitude offset of about 0.3 magnitudes in our sample could produce distance offsets of more than a kpc and the reader may be better served by assuming a constant or color-dependent absolute magnitude of their own choosing.

We have rerun the entire procedure of this paper with the assumption of a 0.054 mas parallax offset folded in, and the findings are largely unchanged (for example, the high metallicity BHB sample's anisotropy changes to 0.73 from 0.70 and our low metallicity sample's anisotropy changes to 0.63 from 0.62). The only major difference is that the velocity dispersions in each component drop by about 15 km s$^{-1}$. This is sensible as increasing the parallax will raise the estimated absolute magnitudes, and therefore lower the estimated distances to the stars in our sample. The proper motions will then imply lesser tangential motions, leading to lower velocity dispersions.

\subsubsection{Optimizing the Machine and Reported Errors}

The performance of the machine can be optimized by changing meta-parameters such as the learning rate, and the maximal tree-depth, etc. To find the optimal parameters, we may split our training data into a training set and a validation set. The machine learns from the training set (75\% of the data in each of the three machines), and then its performance with various meta-parameters is analyzed on the validation set (25\% of the data, this split simulates the machine performance on new, unseen data by hiding a portion of the data from the machine while it learns, to minimize ``data leakage''). We then cycle through various ``splits'' of the training data. This is a technique called ``cross-validation.'' We perform a random grid search of meta-parameters to minimize the mean squared error (for metallicity and absolute magnitude) or to maximize precision (purity, i.e. minimizing the number of false-positive contaminants, \emph{not} maximizing the number of true-positives) for the classification problem.

After this optimization of meta-parameters, we use the entire training set to train the XGB machine before predicting values for the complete observational data set.

The machine self-reports a classification purity (precision) of 85.8\% for BHB stars, 14.2\% contamination, and successfully finds (recalls) 86.2\% of all the BHBs for which there are spectra. We have constructed a confusion matrix based on the results of 100 train-test splits of our training data in Table \ref{tab:confusion_mat}.

\begin{table}
\caption{}
\begin{center}
  \begin{tabular}{ | c | c | c | }
    \hline
    & Precision & Recall \\
    \hline
    BHB & 85.8\% & 86.2\% \\
    \hline
    Not BHB & 90.0\% & 89.8\% \\
    \hline
  \end{tabular}
  \tablecomments{
    The confusion matrix for our XGB classification. Precision (also called purity) indicates the number of true positives as a percentage of all positive classifications. So 85.8\% of all objects classified as a ``BHB'' are true BHBs. Recall (also called completeness) indicates the fraction of objects which are correctly classified. So 86.2\% of all BHBs in our data are classified as BHBs. This table is based on 100 training-testing splits of the data.
}
\label{tab:confusion_mat}
\end{center}
\end{table}

The metallicities are estimated with mean squared error of $\sim$0.15 dex, or $\sigma_{[Fe/H]}\sim$0.39 dex. We investigate the metallicity precision as a function of metallicity in Table \ref{tab:met_prec} by again performing 100 train-test splits of our data. We find that the metallicity error is lower for the bins where training data is abundant, from -2.5 $\leq$ [Fe/H] $\leq$ -1.5, and less precise where training data is sparse.

\begin{table}
\caption{}
\begin{center}
  \begin{tabular}{ | c | c | c | }
    \hline
    [Fe/H] & N & $\sigma$[Fe/H] \\
    \hline
    -3.0 to -2.5 & 43 & 0.77 \\
    \hline
    -2.5 to -2.0 & 169 & 0.31 \\
    \hline
    -2.0 to -1.5 & 232 & 0.23\\
    \hline
    -1.5 to -1.0 & 83 & 0.45 \\
    \hline
    -1.0 to -0.5 & 24 & 0.82 \\
    \hline
  \end{tabular}
  \tablecomments{
      The precision of the metallicity estimate as a function of metallicity. To construct this, we trained the XGB machine with 100 training-testing splits of the total training data and evaluated the precision of the metallicity estimate of the machine on those splits. The precision seems to worsen for the highest and lowest metallicity BHBs, perhaps because of the lower number of objects in those training sets. One BHB from the training set is missing from this table, it has a metallicity of -0.31.
    }
\label{tab:met_prec}
\end{center}
\end{table}

We note that our training data are extremely sparse for high metallicities. Only 25 have metallicities $\geq$ -1 dex. This will cause difficulty for the machine when it comes to predicting metallicities higher than -1 dex.

Absolute magnitudes have precisions around $\sigma_{G}\sim0.41$ and they show no coherent trend with metallicity.

\subsection{Verifying the Regressions}\label{sec:verify}

Through cross-validation, we have obtained estimates for the mean-squared error of the metallicity and absolute magnitude estimates as well as purity estimates for the BHB classification. Here we attempt to verify these in a few independent ways.

\subsubsection{Color-Magnitude Diagram}

A simple way to look at contamination would be to construct a color-magnitude diagram of classified BHB stars. A pure sample should follow the horizontal branch, and contamination should appear off of that branch, most likely at redder and dimmer regions, where the most numerous contaminants (MSA stars) reside.

\begin{figure}
	\includegraphics[width=\linewidth]{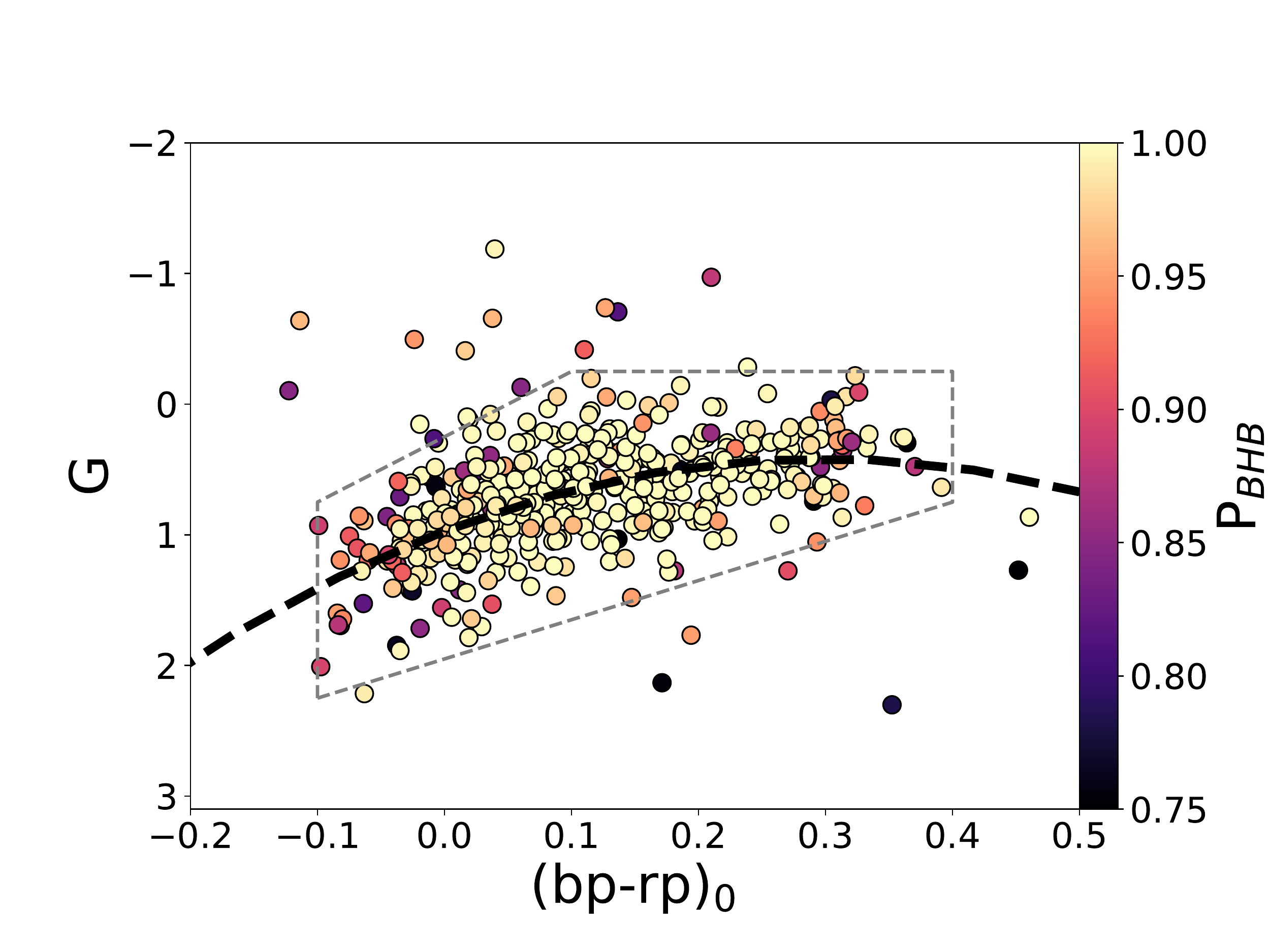}
	\caption{
	  The absolute magnitude of BHB classified stars calculated from parallax as a function of dereddened Gaia color. The stars in this figure are most likely BHBs from their spectra (P$_{BHB}$ $\geq$ 75\%), have less than 10\% parallax error, and are observed at $z>$ 750 pc. Most likely they are all BHBs outside of the plane, so they have full column reddening and a relation between the intrinsic color and magnitude may be constructed (the dashed line is a quadratic fit to the data inside the polygon, which is drawn by eye).\protect\footnotemark[15]
          This figure may be compared to Figure 4 in \citet{dea2011}, which shows a similar BHB color-magnitude distribution, with a similar brightness around 0.6 in SDSS $g$ (our sample of 13,693 BHBs has a mean abs($G$) of $\sim$0.65).
	}
	\label{fig:color_mag}
\end{figure}

In Figure \ref{fig:color_mag} we show the color magnitude diagram for stars estimated to be BHBs by our classifier. The stars are selected only to have $\varpi$ errors of 10\% or less and also to be outside of the plane (z$_{XGB} > $750 pc, so that the color and magnitude values are reliable after extinction correction assuming the full columns of \citealt{sch1998}). We believe that this sample should reasonably reflect the full sample in proportions of BHB and non-BHB stars by referencing Table \ref{tab:training_set} and noting that the proportion of BHBs in the 10\% parallax error sample is 34\%, instead of 42\% in the entire sample. This figure shows a strong concentration of presumptive BHBs, with some contamination extending redwards and fainter (presumably, MSA stars, which includes blue stragglers here) and some contamination brighter and blueward (this could be quasars, extreme horizontal branch stars, or possibly true BHB stars which have been over-corrected for extinction, moving them to too-bright and too-blue sections of the diagram). Traditionally, main sequence A stars are the most prominent contaminants in this type of classification problem.

The selection of height from the plane has been used instead of latitude in an attempt to improve the accuracy of the dereddening procedure (by guaranteeing the full column density of the maps of \citealt{sch1998} are meaningful for each star) while preserving as many stars as possible. However, despite these efforts, areas of low latitude could still be confused by high extinction values, particularly if the extinction maps have poor angular resolution of the dust near our stars. In the sample detailed above: the lowest latitude BHB candidate is at 8.9$^{\circ}$; 97\% of the candidates are above a latitude of 15$^{\circ}$; 88\% are above 20$^{\circ}$; and 68\% above 30$^{\circ}$. There may be some candidates which experience extinction confusion, but the large majority should be in low-extinction directions and outside of the plane with reliable colors and magnitudes. Changing our procedure to include a cut such that all stars are more than 30$^{\circ}$ from the plane does not change the results much, the anisotropy, for example, remains 0.62 for the low metallicity BHBs, and increases to 0.74 from 0.70 for the high metallicity BHBs.

These stars are generally found in the apparent $G$ magnitude range of 10 to 14, with the median near 12.8. Our biggest fear is that some of these objects have been corrected for the full column density extinction of \citet{sch1998}, but in fact experience a different level of extinction. Since these reddening maps are calculated based on extragalactic sources, we expect that the reddening values may over-correct (under-correction should not be a problem), shifting the objects to too-blue and too-bright in this figure (Figure \ref{fig:color_mag}) of the extinction corrected colors and magnitudes. We would expect this type of error to manifest as a diagonal cloud with a negative slope in the figure as main-sequence stars (which should truly be in the bottom-right corner) are dragged toward the center, and BHB stars (which should truly be in the center) are dragged toward the top-left corner. While some of this may be occurring, such a pattern is not prominent or noticeable in this data subset. There also appears to be no dependence on galactic latitude, supporting our choice of using distance from the plane as a selector.

Note that Figure \ref{fig:color_mag} has heavier contamination when we include stars closer to the plane and of lower probability of being a BHB star than in our shown plot, which has cuts at 750 pc and 75\%, respectively. Including lower probability stars increases contamination in the dimmer-and-redder direction, most likely MSA stars. Including stars closer to the plane populates a plume of high probability BHB ``contaminants'' which are bluer and brighter than the BHB locus. This could be true BHB stars which are extinction corrected more than they are intrinsically dimmed by dust.

To get a rough estimate of the contamination levels, we draw a polygon around what we presume to be ``true'' BHBs by eye. 23 out of 450, or 5\% of the plotted stars, lie outside this polygon. This is lower than the machine-reported 14\% contamination, although, there may also be contamination inside of this selection box. If we construct the same figure with cuts of 50\% P$_{BHB}$ and z $>$ 500 pc, this changes to 10\% lying outside of the polygon. This color-magnitude diagram is indicative of our data in this current work, but not for the entire catalog. The interested party is encouraged to take note of this. We discuss the contamination profile inside the plane more thoroughly in a companion paper.

\footnotetext{Including lower probability BHB stars (P$_{BHB}$ between 0.5 and 0.75) increases main sequence contamination to the red and dim corner of the plot. Including lower $z$ height data creates a plume of contamination along the reddening vector to the brighter, bluer corner, which could be true BHBs incorrectly dereddened. This figure should be simple to reproduce with the provided catalog crossmatched to Gaia, so the interested reader is encouraged to examine how these effects change depending on your own cuts.}

\subsubsection{Globular Clusters}\label{sec:globs}
\begin{figure*}
	\includegraphics[width=\linewidth]{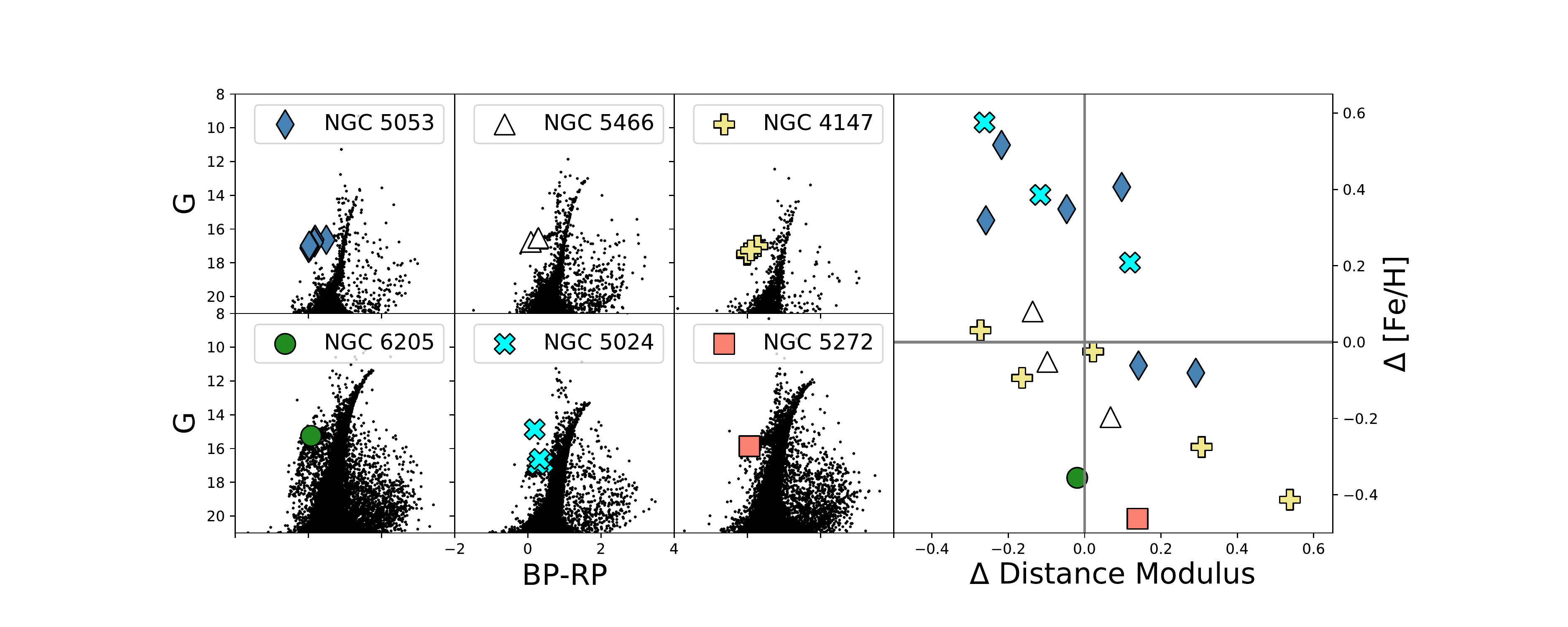}
	\caption{
          Color magnitude diagrams of six globular clusters from the \citet{har1996} catalog which have BHB stars within 1 tidal radius of their center in our data. The \emph{right} panel shows the difference between the individual stars [Fe/H] and distance moduli in relation to the catalog values for their respective clusters.\protect\footnotemark[16] \\
          There is an apparent covariance in XGB predicted G magnitudes and [Fe/H] values where stars which are too metal poor are predicted to be too bright, and stars which are too metal rich are predicted to be too dim. The [Fe/H] and $G$ values are predicted by independent machines.
        }
	\label{fig:glob_bhb}
\end{figure*}

Globular clusters have well defined distances and metallicities in the literature. By cross-matching our data on the sky within 1 tidal radius of the globular clusters in the Harris Catalog (\citealt{har1996}; 2010 revision), we find 20 stars classified as BHBs in six clusters. We show the distance modulus and abundance values of these six clusters and 20 stars in Table \ref{tab:glob_bhb} and plot them in Figure \ref{fig:glob_bhb}.

With regard to the apparent visual outlier in NGC 5024, we find that the three stars lying visually on the horizontal branch have proper motions of ($\mu_{R.A.}$, $\mu_{Dec.}$) = (-0.09, -1.64), (-0.25, -1.9), (-0.14, -1.26), and the outlier has a proper motion of (-0.31, -1.42). This proper motion seems consistent with cluster membership (as does the radial velocity). It could possibly be a true BHB member of the cluster with a highly uncertain photometric magnitude, although that seems unlikely as its flux error is well below 1\%. It is also not likely to be an RR Lyrae (which would be located in a similar place on the color-magnitude diagram), as it is observed two magnitudes from the horizontal branch when RR Lyrae typically only have pulsations on the order of one magnitude. This star is removed from the subsequent analysis of the [Fe/H] and abs($G$) errors, although, we note that including it actually improves the expected precision of the [Fe/H] calculation.

\begin{table*}
\caption{}
\begin{center}
\begin{tabular}{|c|c|c|c|c|c|}
\tableline
Globular & Glob. [Fe/H] & BHB [Fe/H] & Glob D. Mod. & BHB D. Mod. \\
\tableline
NGC 6205  &  -1.53  &  -1.89 &  14.26  &  14.24 \\
\hline
NGC 5466  &  -1.98  &  -2.03 &  16.02  &  15.92 \\
 &  &  -2.18 &  &  16.09 \\
 &  &  -1.9 &  &  15.88 \\
\hline
NGC 5272  &  -1.5  &  -1.96 &  15.04  &  15.18 \\
\hline
NGC 5053  &  -2.27  &  -1.75 &  16.2  &  15.99 \\
 &  &  -1.95 &  &  15.94 \\
 &  &  -1.92 &  &  16.16 \\
 &  &  -2.35 &  &  16.49 \\
 &  &  -2.33 &  &  16.34 \\
 &  &  -1.86 &  &  16.3 \\
\hline
NGC 5024  &  -2.1  &  -1.96* &  16.26  &  14.27* \\
 &  &  -1.89 &  &  16.38 \\
 &  &  -1.71 &  &  16.15 \\
 &  &  -1.52 &  &  16.0 \\
\hline
NGC 4147  &  -1.8  &  -1.77 &  16.43  &  16.15 \\
 &  &  -2.21 &  &  16.97 \\
 &  &  -1.89 &  &  16.26 \\
 &  &  -2.07 &  &  16.73 \\
 &  &  -1.82 &  &  16.45 \\
\hline
\end{tabular}
\tablecomments{
  BHBs which reside within 1 tidal radius of the globular clusters shown in Figure \ref{fig:glob_bhb}. The asterisks indicate an item omitted from the right panel of Figure \ref{fig:glob_bhb} because its distance modulus is a large outlier ($\sim$2 mag).
}
\label{tab:glob_bhb}
\end{center}
\end{table*}

We compare the estimated [Fe/H] values of the member BHBs from the XGB regression with the literature values of their clusters' metallicities from the Harris Catalog. The error (assuming that the catalog values are exact and true) for our 19 stars (20, minus one visual outlier) is $\sigma_{[Fe/H]}\sim0.31$. This is slightly more precise than the machine estimated value of $\sigma_{[Fe/H]}\sim0.39$. Perhaps this is because the metallicities of these globular clusters lie between -2.5 and -1.5, where we predict the [Fe/H] errors to be lower (see Table \ref{tab:met_prec}).

For each star in these globular clusters, we may also assign a distance modulus value,

$$mod_{XGB} = G_{XGB}-g_{0},$$

which can be compared to the distance moduli of the globular clusters, mod$_{G.C.}$.

If we assume that the error on the observed magnitude, g$_{0}$, and the distance modulus of the globular cluster, mod$_{G.C.}$ are small compared to the error on G$_{XGB}$, then:

$$\sigma(mod_{XGB} - mod_{G.C.}) \sim \sigma_{G_{XGB}} \sim 0.21. $$

\footnotetext{The obvious visual outlier in NGC 5024 is omitted from the right-hand panel as its distance modulus is an extreme outlier. The star is presented in Table \ref{tab:glob_bhb}.}

This value of $\sigma_{G}\sim0.21$ is substantially lower than the error estimated by the machine of $\sigma_{G}\sim0.41$. When the outlier discussed previously is included, $\sigma_{G}\sim0.48$. While we don't have a definitive reason as to why, it could be that our data used for training the absolute magnitudes has some MSA or blue straggler star contamination which increases the magnitude estimate spread, while the globular cluster fields, being old populations, are deficient in this contamination.

Note that the purity of this sample is about 95\% (if the star lying off the horizontal branch, in NGC 5024, is in fact a contaminant and not a field BHB). This is similar to the machine-reported contamination values, but should not be given too much weight as these observational fields should naturally be over-abundant in BHBs.

In Figure \ref{fig:glob_bhb} we note that there seems to be a slight covariance between the distance modulus offset and the metallicity offset. To try and correct for this, we attempt a multi-output regression, that is, fitting the metallicity and the absolute magnitude simultaneously. Unfortunately, the Century Survey combined with the \citet{xue2008} sample only have 9 BHBs with metallicities and good parallax measurements which lie outside the plane (having good $G$ measurements). These 9 objects also only span -2.0 $\leq$ [Fe/H] $\leq$ -1.5, which would exacerbate our noted problems with predicting high metallicity values for our data. We abandon this line of inquiry and accept the covariance.

This covariance could be an effect of how temperature and metallicity affect spectral line shape. Larger brightnesses tend to characterize lower surface gravity stars, which have narrower spectral lines. Larger abundances tend to make lines deeper. A deep-and-wide line (high abundance, low brightness) could have a similar overall shape to a shallow-and-narrow line (low abundance, high brightness) on a normalized spectra. Confusion between these two could cause the machines to mistakenly lower the abundance of a given spectrum while also increasing the brightness, leading to the covariance we see.

This could possibly be avoided with higher resolution spectra as these two effects actually affect the line shape in subtly different ways, particularly in the wings of the profile.

\subsubsection{Duplicates}
\begin{figure}
	\includegraphics[width=\linewidth]{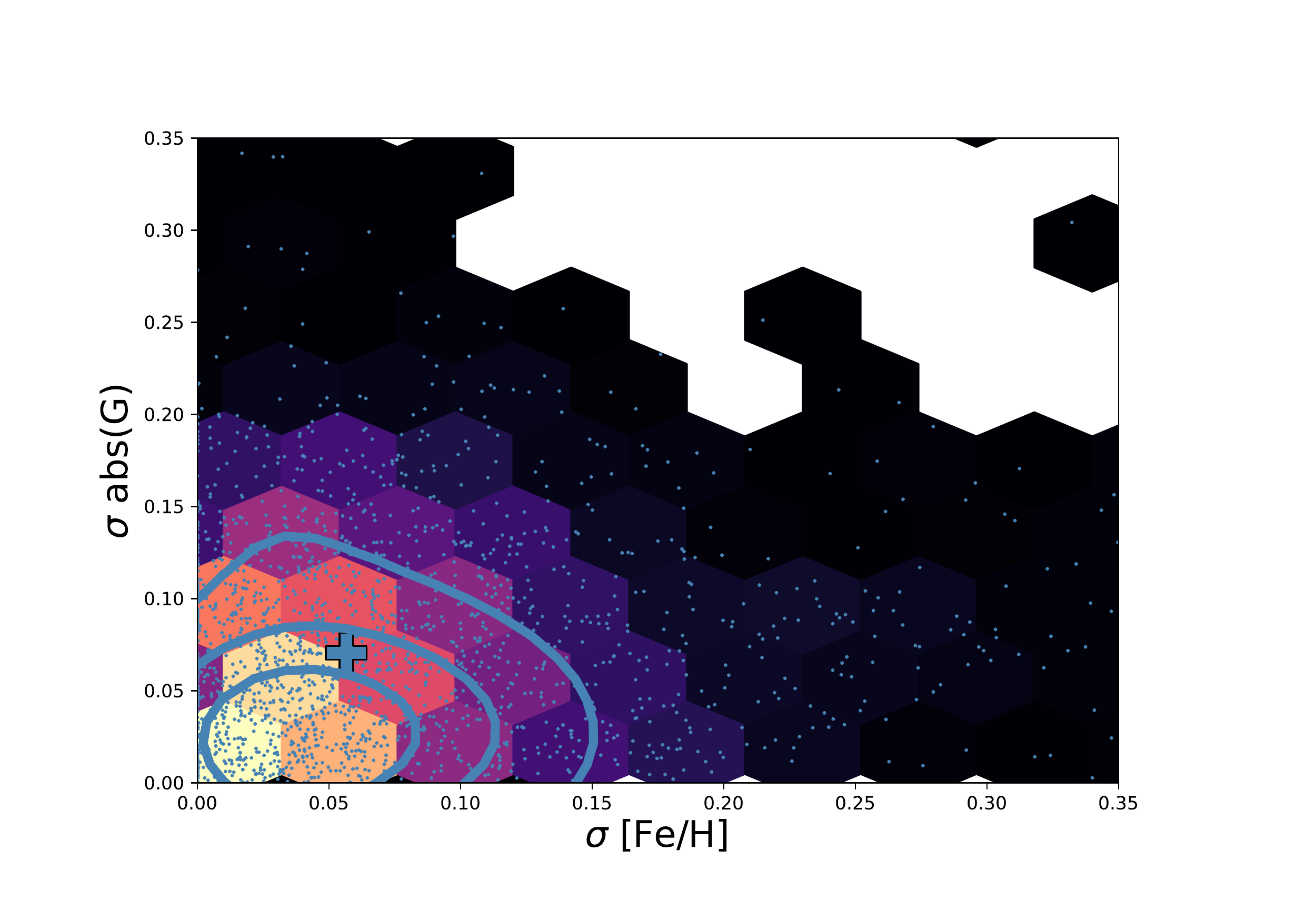}
	\caption{
          The $\sigma[Fe/H]$ and $\sigma G$ values for stars (blue points and hex map) observed two or more times (with different observing conditions) by LAMOST. The contours indicate the smoothed 25\%, 50\%, and 75\% confidence intervals and the cross indicates the median $\sigma[Fe/H]$ and $\sigma G$ (0.07 and 0.06, respectively).
        }
	\label{fig:dupes}
\end{figure}

The LAMOST survey has observed numerous stars in multiple observations, some of these duplicate observations have been classified by the XGB machine as BHB stars. Since the spectra are independent of each other, they provide some insight into the internal errors of the classifications and regressions. In Figure \ref{fig:dupes} we show the distribution of magnitudes and metallicities for stars observed two or more times. From this Figure, we see that the median $\sigma [Fe/H]$ is about 0.07 and the median $\sigma G$ is about 0.06\footnote{These values are much smaller than the dispersions discussed previously, that is, 100 spectra of similar metallicity or brightness stars would have a larger spread in [Fe/H] and abs(G) than 100 spectra of the same star. This may imply that the fitting surface is not smoothly varying, which is a side effect of random forest algorithms, or that some third parameter such as surface gravity or temperature works to inflate the errors.}.

Note that duplicates are removed in the analysis section of the paper (preserving the observation with the highest signal to noise ratio) but are not removed from the catalog.

In our sample of 13,693 BHB spectra, we estimate that there are 11,046 unique stars. There are 1,884 objects with multiple-observations, the largest number of repeated observations is one star with 16 individual spectra and the median is two observations per star.

\section{Coordinates and Velocities}\label{sec:coords}
Kinematic phase information is calculated in the usual way from observed coordinates, proper motions, radial velocities, and the distances derived from the XGB predicted absolute magnitudes and photometry which has been extinction corrected using the full column dust maps of \citet{sch1998}\footnote{We use full column reddening here as we will later cut the data to have a distance of more than 3 kpc from the plane, to select halo stars. When using these data at lower altitudes, a different procedure is necessary.}.

Our coordinates are all right handed with the Sun at $(X, Y, Z) = (-8.27, 0, 0)$, rotating with a velocity of $(V_{R}, V_{\phi}, V_{Z}) = (0, -236, 0)$ km s$^{-1}$. $V_{R}$ is positive away from the Galactic center. The solar motion is $(U, V, W) = (13.0, 12.24, 7.24)$ km s$^{-1}$ with respect to the local standard of rest. These positions and velocities are all adopted from \citet{sch2017, sch2012}.

The final assignment of these coordinates is based on the mean value of 100 realizations of the observed star considering the observational errors as well as the Gaia correlations on the relevant parameters.

\subsection{Final Cuts for Halo BHBs}\label{sec:qual}

\begin{figure}
	\includegraphics[width=\linewidth]{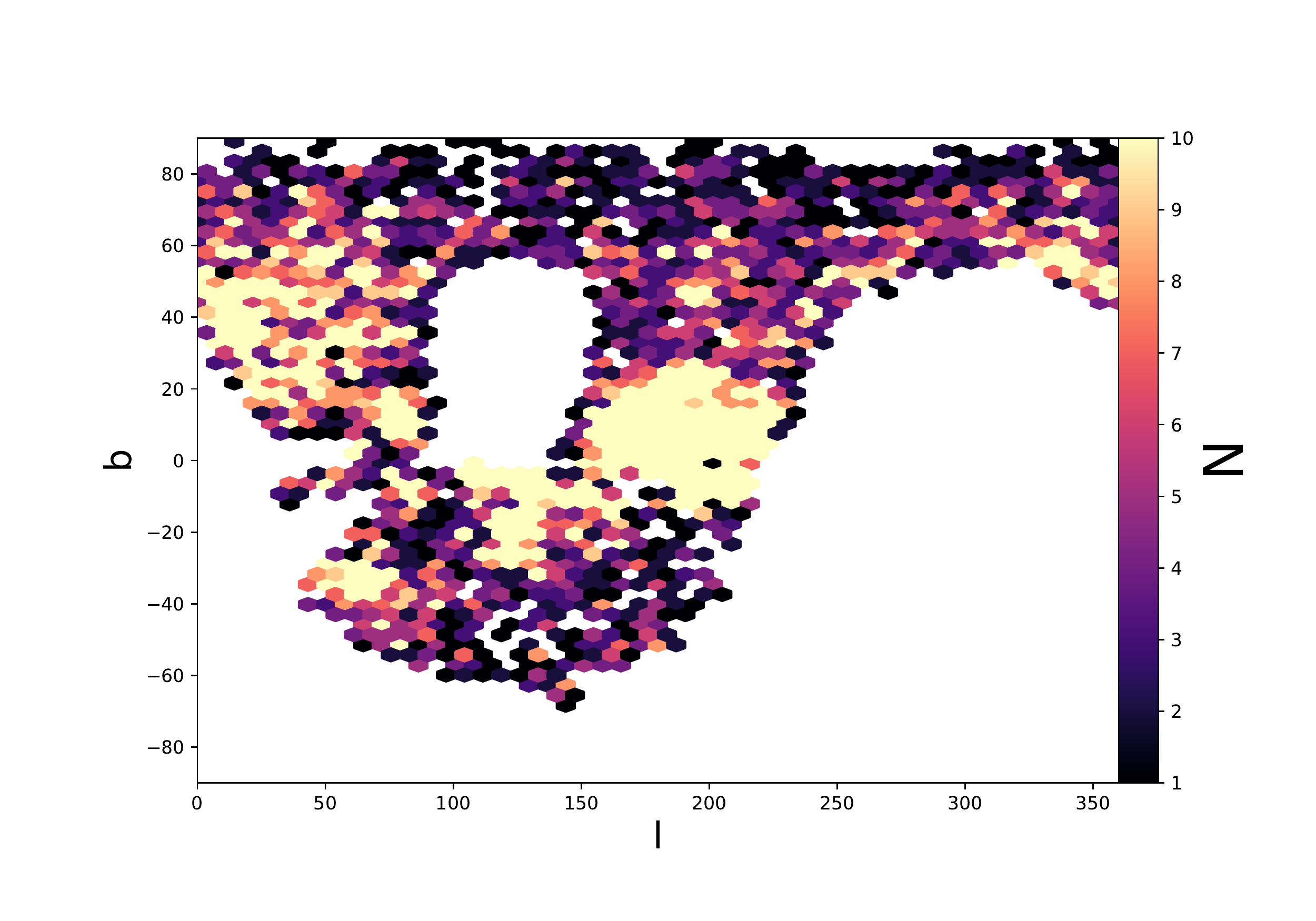}
	\caption{
		Sky coverage of our BHB sample. It can be seen that our BHB sample covers a good amount of the plane which is not covered in other large BHB catalogs. Note that this is the full catalog, the subset analyzed here is restricted to more than 3 kpc away from the plane.
	}
	\label{fig:footprint}
\end{figure}

Coverage of our sample is shown in Figure \ref{fig:footprint}. The catalog covers a large area of sky unexplored by prior BHB surveys. However, for the rest of this paper, we will only consider stars which are observed at more than 3 kpc from the plane, to select halo BHBs. In a companion paper we will more thoroughly analyze the BHB population closer to, and inside of, the plane.

From our initial Gaia-LAMOST crossmatch of 155,549 objects, 14,101 are classified as BHBs. From the catalog we remove objects for which the machine has predicted values outside of the input parameter range for G (from 2.5 to -1.2) and [Fe/H] (from -3 to -0.3), which leaves 13,693 BHBs in our final catalog.

We make the following additional cuts to the catalog to prepare the scientific sample for the remainder of this paper: duplicates removed within 7'', e$\mu_{R.A.}$, e$\mu_{Dec.}$ $<$ 10\%, e(r.v.) $<$ 10\%, P$_{BHB}$ $>$ 75\%, $z$ $>$ 3 kpc, 3$\sigma$ velocity outliers iteratively culled.

This leaves 2,692 high confidence BHBs in our halo sample with good kinematics which form the data for the rest of the current analysis.

\section{Analysis}\label{sec:analysis}

\subsection{Anisotropy}

Anisotropy is an observable measurement of the degree to which a population is orbiting tangentially versus radially in the galaxy:

\begin{equation}\label{eqn:anisotropy}
\beta = 1-(\sigma_{\theta}^{2} + \sigma_{\phi}^{2})/(2\sigma_{r}^{2}).
\end{equation}

Populations which are radially biased have $\beta > 0$, and tangentially biased populations have $\beta < 0$.

\begin{figure*}
	\includegraphics[width=\linewidth]{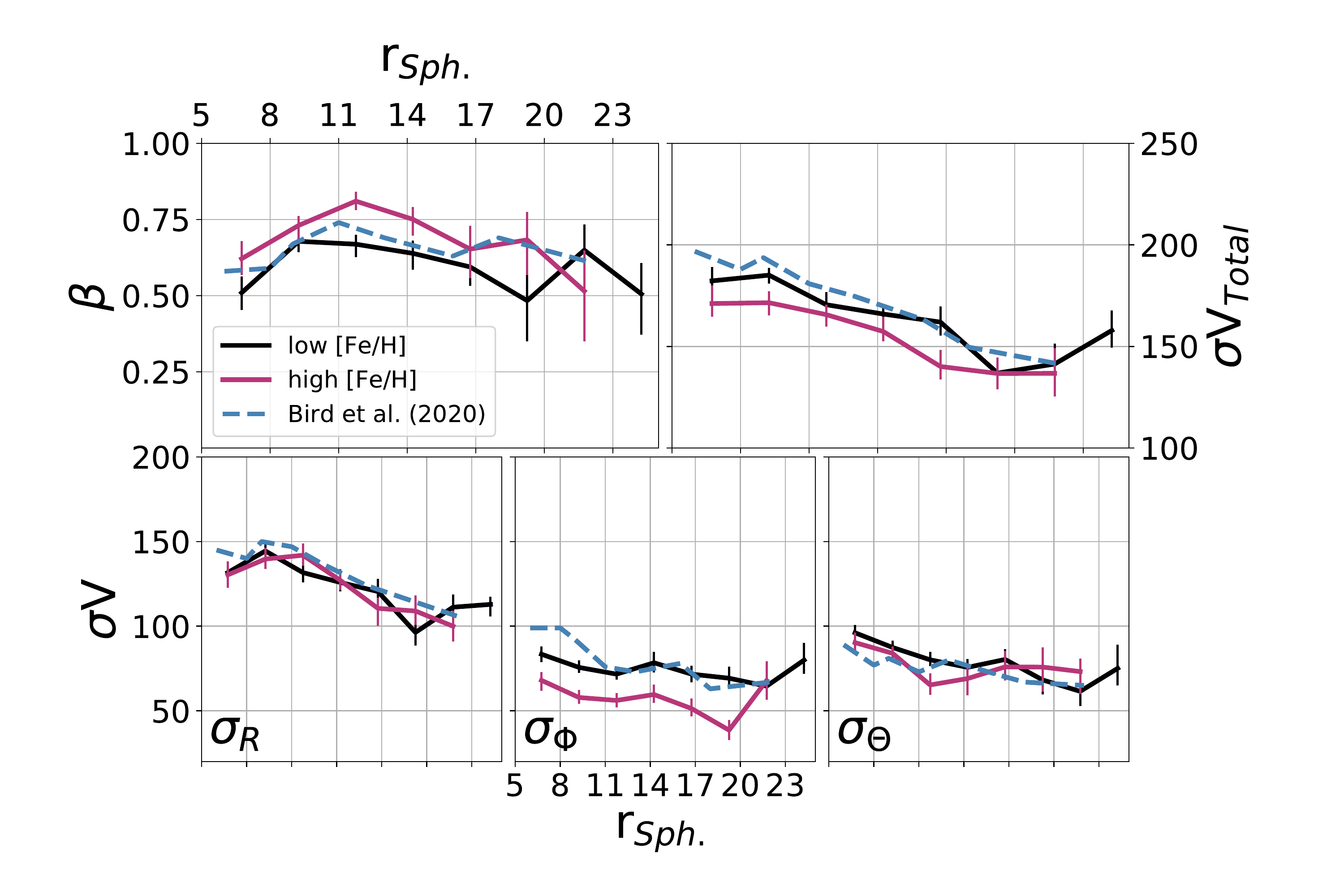}
	\caption{
		\emph{Top Left:} anisotropy is a measure of how radial the orbits of a population are. The metal rich stars in our sample are slightly more radial than our metal poor stars.
		\emph{Top Right:} total velocity dispersion of the BHB sample falls with spherical radius, the metal rich BHBs usually being cooler than the metal poor BHBs.
		\emph{Bottom:} Individual components of the velocity dispersions, used to derive the upper panels. \\
                The panels include the findings of \citet{bir2020} for their ``all SDSS halo BHB'' sample, and the agreement seems good. We refer the reader to that paper for a more detailed analysis of their BHB and K-giant data including metallicity dependence and a more thorough removal of substructure than we have implemented here. We have used the ``extreme deconvolution'' method of \citet{bov2011} to calculate the velocity dispersions.
	}
	\label{fig:aniso_disp}
\end{figure*}

In Figure \ref{fig:aniso_disp} we plot the velocity dispersions and anisotropy of our sample in high and low metallicity subsets as a function of spherical radius. We find that the halo sample has relatively flat anisotropy for both high and low metallicity bins, with the low metallicity stars being more tangential velocity dispersion dominated ($\beta \sim 0.62$) and the high metallicity stars being more radial velocity dispersion dominated ($\beta \sim 0.70$). We note that the radial velocity dispersion is similar between the two populations, the tangential velocity dispersion of the metal rich stars is lower than that of the metal poor stars.

\citet{bir2019} analyzed the anisotropy parameter using LAMOST K-giants, which are a younger population than BHB stars. They found a generally more radial population, with $\beta \sim$0.8 at Galactocentric radii less than 20 kpc, which becomes more tangential at larger radii before reaching an almost isothermal value of 0 at 100 kpc. Interestingly, they also split their population based on metallicity (their figure 5) and found that their lowest metallicity K-giants have $\beta \sim$0.6 while higher metallicity K-giants are closer to $\beta \sim$0.8. This metallicity trend, of larger $\beta$ values for metal rich stars, is consistent with our results, although our population of BHB stars seems to be generally more tangential velocity dispersion dominated than their population of K-giants.

\citet{weg2019} recently looked at the anisotropy and velocity ellipsoids of RR Lyrae stars out to a spherical Galactocentric radius of about 20 kpc. Their sample is similar in both age and extent to our current BHB population. They found that the velocity profiles of RR Lyrae become highly radial beyond 5 kpc with $\beta \sim$0.8; inside that radius, they are more tangential, with a prograde rotation\footnote{Our sample is slightly retrograde in the area studied ($z$ $>$ 3 kpc), near -20 km s$^{-1}$.} of about 50 km s$^{-1}$ and $\beta$ down to 0.25. Our sample is less radial velocity dispersion dominated than their sample, and does not probe as low in galactocentric radii as theirs does. It's worth noting that their study probed latitudes down to 10$^{\circ}$, which would bring their observational window closer to the disk than the one considered in this work.

Analyses of cosmological simulations generally find the $\beta$ parameter to rise with galactic radius. At their centers, simulated galaxies are nearly isotropic. Moving outward, the halos of galaxies tend to become more radially biased, with $\beta$
rising, at first quickly and then more slowly, to a value $\geq$0.7 past 10 kpc from the center (see \citealt{loe2018} and discussion therein).

We note that we have not removed substructure in this analysis. Our sample should not suffer from large amounts of major substructure contamination, as our sample does not extend to the distance of major contaminants such as Sagittarius (our sample cuts off around 20 kpc).

\subsection{Ages and Metallicities}\label{sec:age_met}

Nearly thirty years ago, \citet{pre1991} noted that BHB stars exhibited a color gradient in the halo, growing redder with increasing radius. They posited that such a color gradient could be the result of an age gradient in the halo, with redder BHB stars being, on average younger\footnote{They noted that the color gradient could also be the result of more distant BHB stars having smaller core masses, but found no reasonable explanation for why that may be the case; see also ``the second parameter phenomenon'' of BHB morphology, explored in detail in \citet{dot2010}.}.

That work has been revisited recently using Sloan Digital Sky Survey spectroscopy \citep{san2015} and photometry (\citealt{car2016}, see also \citealt{whi2019}). They similarly find a suggestion of a reddening of BHB color with increasing radius, with bluest (oldest) stars being more centrally concentrated.

\begin{figure*}
	\includegraphics[width=\linewidth]{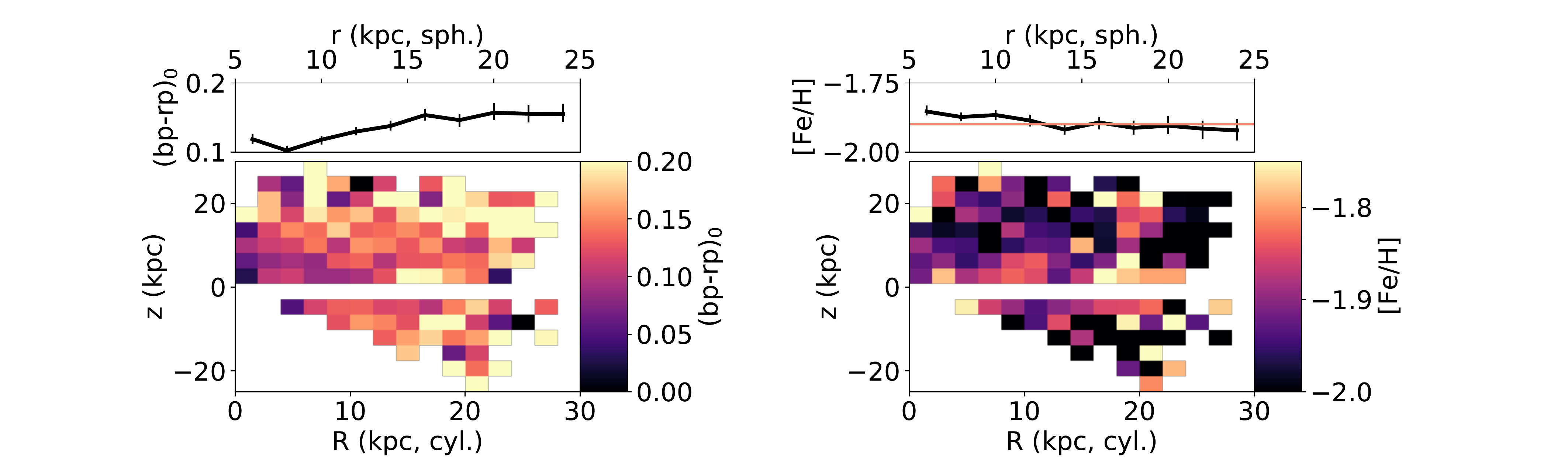}
	\caption{
          \emph{Left}: The colors of our BHB stars as a function of distance from the plane and radius (cylindrical in the bottom frame, spherical in the upper frame). Color may be a proxy for age in BHBs, with redder BHBs coming from younger populations. This figure shows a clear trend for more distant BHB stars to be redder, and therefore to possibly represent a younger population. \\
          \emph{Right}: Similar to the lefthand frames, but with metallicity instead of color.
        }
	\label{fig:age_met}
\end{figure*}

We plot the colors, as well as the metallicities of our BHB sample in Figure \ref{fig:age_met}. This Figure shows that, as we move outward in the halo, the BHB population grows redder (perhaps younger) and metallicity remains relatively flat around -1.9 dex. This abundance value is similar to what is expected for the halo, if a bit high, \citet{xue2008} found a value of around -2 dex in their sample.

\section{Discussion}\label{sec:discuss}

In our halo sample, we have noted two trends:
\begin{itemize}
  \item Our metal rich BHB stars move on more radial orbits than our metal poor stars at all radii.
  \item As we move outward in the halo, the BHBs grow redder (possibly younger) while metallicity remains relatively flat.
\end{itemize}

When speaking of the halo, there are a few main paradigms. The first is the idea of an inner-outer halo, two major components which overlap but trade dominance as radius increases leading to a sort of ``break'' where an outer halo becomes the more dominant population. The inner halo is thought to be slightly prograde and more metal rich, while the outer halo is thought to be slightly retrograde and more metal poor. This idea has been pursued vigorously with mounting evidence from the research group of \citet{car2007}. This finding has been supplemented with evidence from other teams revealing broken halo density profiles in various tracers (e.g.: BHBs \citealt{dea2011} and \citealt{dea2018}, RR Lyrae \citealt{ses2011}), in the presence of two major components in the chemistry (\citealt{nis2011}, and subsequent papers in that series), in a broken age profile in the halo \citep{whi2019}, and in the exotic stellar population prevalences \citep{car2012}.

A more recent development, unveiled primarily through the exquisite proper motion data from Gaia, is couched in terms of merger history. The discovery papers of \citet{bel2018} and \citet{bel2020} detail two major halo components dubbed ``The Sausage'' and ``The Splash,'' respectively. The Sausage is a highly radial population of stars thought to have arisen from a merger 8-11 Gyr ago with a 10$^{10}$ M$_{\odot}$ galaxy, and the Splash is thought to be debris from the Milky Way's proto-disk at the time of this merger. The Splash is slightly younger by perhaps a Gyr and more metal rich ([Fe/H] $>$ -0.7, while the Sausage is found around [Fe/H] between -1.7 and -0.7). The Splash stars generally have low angular momentum and some have retrograde orbits (see also \citealt{ama2020}).

The ``Gaia Enceladus'' theory from \citet{hel2018} combines the retrograde halo with the eccentric Sausage into a single event. Other groups (e.g. \citealt{mye2019}) propose that the Sausage arose from a single merger while the retrograde component arose from a separate event called ``Sequoia.'' Seqouia stars are thought to be slightly more metal poor than Sausage stars.

\begin{figure}
	\includegraphics[width=\linewidth]{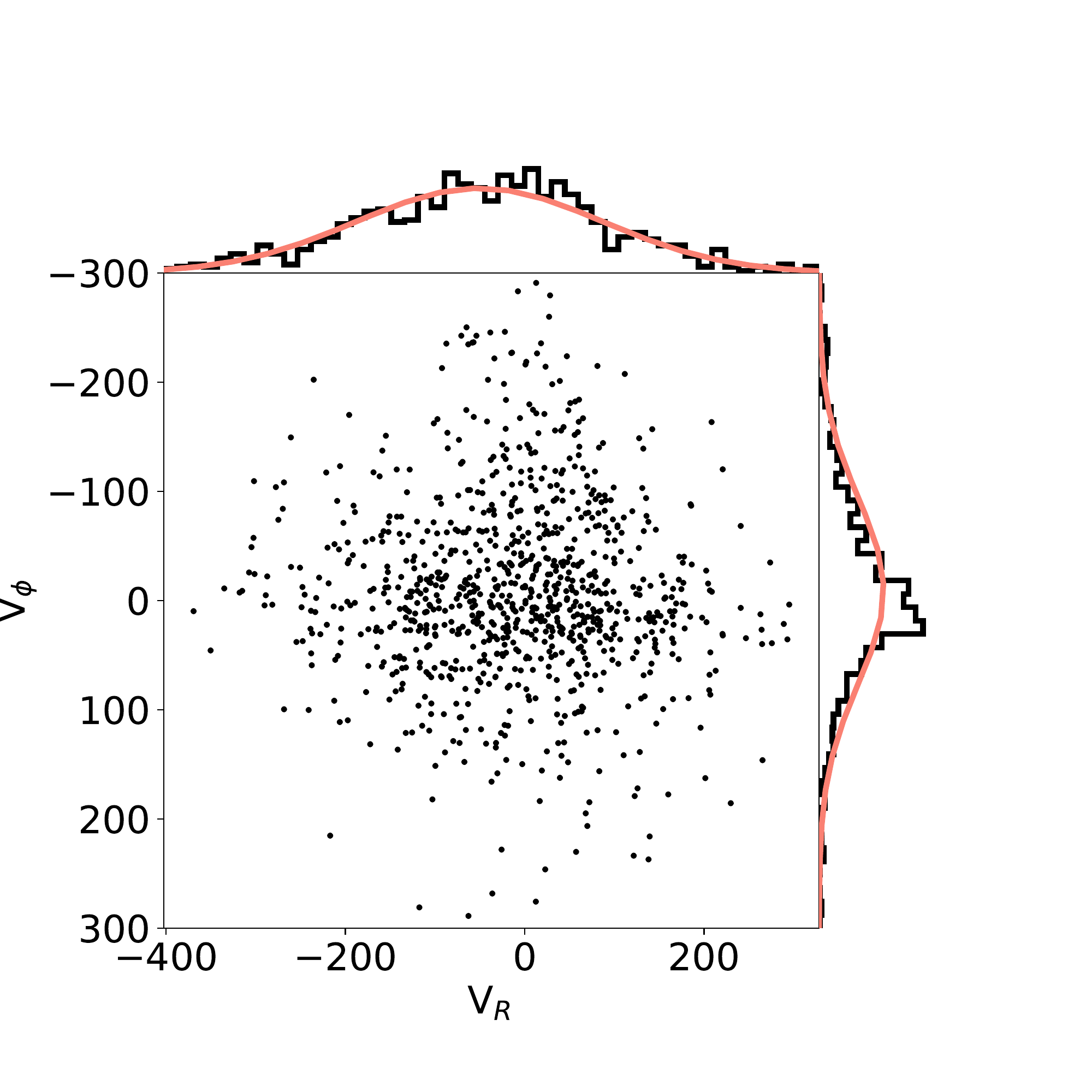}
	\caption{
          The V$_{R}$ V$_{\phi}$ plane of our bhbs with metallicities $>$ -1.8 and $Z$ $>$ 3 kpc. It can be seen that our sample is roughly gaussian in V$_{R}$, as expected for the canonical halo population. However, the V$_{\phi}$ distribution is not well fit by a gaussian, with a strong peak near zero and a heavy tail extending toward prograde rotation. This peak near zero is indicative of the sausage-like feature in \citet{bel2018}, their Figure 2. The heavy tail extending toward prograde rotation could be related to the thick disk halo interface.
        }
	\label{fig:sausage}
\end{figure}

Comparing the metal-rich to metal-poor components at different radii, we find that our metal rich BHBs seem to be more radial velocity dispersion dominated (having higher $\beta$), which could imply that they contain a significant portion of ``Sausage'' members. We briefly inspect the V$_{R}$ - V$_{\phi}$ velocity plane in Figure \ref{fig:sausage}, and find that a portion of our high metallicity stars reside in an extended V$_{R}$ feature around zero rotation, which is consistent with the ``Sausage'' kinematic feature.

We briefly check for the presence of the Sequoia / Enceladus feature by investigating our most metal poor BHBs; however we do not see evidence of these less enriched BHBs having a surplus of counter-rotating members, so we do not think a substantial amount of stars from those features are present in our sample.

At this point, we return to our other finding that our BHB stars possibly grow younger with increasing radius, while the metallicity gradient remains flat.

When looking at the globular cluster population of the Milky Way, in-situ globular clusters generally populate an area around 12-14 Gyr of age with metallicities mainly between -0.5 and -1.5, and a slight tail extending to lower metallicities. Accreted globular clusters follow a track which moves from age = 14 Gyr and [Fe/H] = -2.5 dex to age = 10 Gyr and [Fe/H] = -1.25 dex (that is, a track which is more metal poor for a given age, or younger at a given metallicity, than the in-situ clusters, see \citealt{for2020})\footnote{The Sagittarius associated globular clusters, for example, have ages up to four Gyr younger than the in-situ Milky Way clusters at similar enrichment levels, and the Sagittarius stream is prominent at distances up to 80 kpc in the halo. Alternatively, the Gaia Enceladus associated globular clusters have metallicities 1 dex lower than in situ globular clusters at similar ages.}.

It has also been suggested in simulations that minor mergers, with masses in the range of 1:50 or 1:100 that of the Milky Way, are expected to deposit more debris in the outer regions of the Galaxy compared to major mergers \citep{kar2019}.

In $\Lambda$CDM simulations of Milky Way type galaxies, it is often found that mass buildup is dominated by early accretions \citep{bul2005}. Since these early accretions are generally larger, they experience more dynamical friction and sink toward the center, this leads to the oldest stars being most common in the very central portions of the galaxy \citep{tum2010}. If there are several accretions, it can flatten the halo metallicity gradient, and if there are few, then a steeper gradient is expected \citep{coo2010}.

It is also expected that systems in more quiescent regions may continue star formation longer than those which are quenched through merging processes and chaotic environments \citep{bal1999}. Therefore smaller accretions at later times could possibly host younger populations.

In our data, we see a relatively flat metallicity gradient, which would imply that the Milky Way has experienced several mergers rather than a few. We also see a color gradient, which could be interpreted as an age gradient, with younger stars in the external regions. We suspect that these younger, peripheral stars come from minor mergers with smaller (shifting them to the external regions of the halo) satellite systems which have less intense star formation owing to their lower masses (shifting them to lower metallicities for a given age), and have been accreted at later times than the older halo stars from earlier major mergers (allowing their star formation to continue longer outside of the quenching effects of accretion, and so, host younger stars).

In short, we see evidence in the anisotropy parameter for a portion of our data to be associated with the Gaia Sausage event (having a higher $\beta$ and therefore being more radial). We do not see indications of our data belonging to the Splash or Sequoia/Enceladus as we see no retrograde component beyond what is expected from the canonical halo at low metallicities. We suggest that we see evidence for the outer regions of our footprint being predominantly accreted stars from smaller accretion events at later epochs.

\section{Conclusions}\label{sec:conclusion}
We have implemented a machine learning algorithm to classify LAMOST spectra with blue Gaia colors as BHB or not-BHB objects. This classification is approximately 86\% pure. \emph{Please note when using this catalog that this is more complicated for lower machine-probability stars and stars closer to the plane, although, even in the worst case, the catalog should be at least $\sim$60\% pure}. We will discuss this in more detail in the companion paper. We similarly (with a second machine) predict metallicities to about 0.35 dex, although we note that our sample does not contain many [Fe/H] values above -1 dex. This is probably a limitation of our training data, and it is better to consider the derived metallicities as relative values, instead of absolute values. We predict absolute magnitudes to $\sim$0.31 mag using a third machine which is trained using parallax distances. We verify the classification and regressions through investigation of color magnitude diagrams of the sample and globular cluster comparisons. In this comparison we note that there is covariance between the estimated metallicity and absolute magnitude, with stars being estimated as too bright also being too metal poor. However, despite that, comparison with globular clusters yields errors near 0.3 dex in [Fe/H] and 0.2 in magnitude, with no significant systemic offset from cluster literature values.

This catalog comprises the largest database of BHB stars thought to currently be inside the extent of the disk. Other large spectroscopic catalogs, notably those from the SDSS, largely avoid the plane latitudes and have higher apparent magnitude limits, which forces their BHB observational windows out of the disk regions. Color-based selection of BHBs inside the plane is difficult since color is reddening dependent\footnote{Particularly ultraviolet wavelengths such as the U band, which is frequently used for BHB identification.} and reddening is largely unknown inside the plane, with the exception of three-dimensional reddening maps which remain difficult to construct precisely\footnote{Despite outstanding efforts from, for example \citet{dri2003}, \citet{mar2006}, \citet{sal2014}, \citet{gre2019}.}. Outside the plane, extinction may be regarded as a known quantity thanks to the invaluable maps of \citet{sch1998}. Spectroscopic identification, as we have performed here, does not suffer from this extinction confusion though. This catalog therefore presents interesting and novel opportunities to study this species of star in a hitherto unexamined regime.

We have briefly investigated the halo properties of our BHB stars. We find that our metal rich BHBs are on more radial orbits at all galactocentric radii. We interpret this as showing that the metal rich population in our sample is populated by Gaia Sausage stars in addition to halo stars, while the metal poor stars do not obviously belong to any of the other named major merger components.

We find that as we move outward in the halo, from about 5-20 kpc, the BHBs grow redder (which could mean younger) and that the metallicity gradient remains mostly flat.

If we suppose that the inner region is populated mostly by large accretion events, and the outer region by smaller ones, then we would expect a decreasing metallicity gradient and a flat age gradient if the mergers occurred at the same time and star formation in the systems was subsequently quenched simultaneously (since larger systems have higher star formation rates and should be intuitively more enriched).

If, instead, the larger, centrally concentrated accretions occur before the smaller accretions, their star formation will be quenched at earlier epochs, so we would perhaps see a decreasing age gradient in the stellar populations as we move outward in the halo to the areas populated by smaller, more recent mergers. In this situation, we may not expect a decreasing metallicity gradient with radius; since the outer regions are quenched at later epochs, they may have had more time to enrich and their abundance levels may ``catch up'' to those of the larger systems at the center.

In this paper, we have investigated only BHB stars in our catalog which reside far from the plane, in the halo, omitting more than half of our data. Dealing with the data in the plane requires a more nuanced and rigorous investigation of effects from differential reddening (since our distances here are photometrically derived) and larger contaminating populations. A second paper has been prepared which deals specifically with these data inside the plane.

The catalog may be downloaded from https://zenodo.org/record/4547803

\section{Acknowledgements}

We thank the referee for their thoughtful comments which helped improve the clarity of this paper.

We thank Corrado Boeche for his work in normalizing the LAMOST spectra so that we could easily utilize them in this work.

We thank Iulia Simion for helpful discussions and suggestions regarding Gaia kinematics.

We thank the developers and maintainers of the following software libraries which were used in this work: Topcat \citep{tay2005},  NumPy \citep{van2011}, SciPy \citep{jon2001}, AstroPy \citep{ast2013}, matplotlib  \citep{hun2007}, scikit-learn \citep{ped2011}, IPython \citep{per2007}, XGBoost \citep{che2016} and Python.

The research presented here is partially supported by the National Key R\&D Program of China under grant No. 2018YFA0404501; by the National Natural Science Foundation of China under grant Nos. 12025302, 11773052, 11761131016; by the ``111'' Project of the Ministry of Education under grant No. B20019. This work made use of the Gravity Supercomputer at the Department of Astronomy, Shanghai Jiao Tong University, and the facilities of the Center for High Performance Computing at Shanghai Astronomical Observatory.

JJV gratefully acknowledges the support of the Chinese Academy of Sciences President's International Fellowship Initiative. M.C.S. acknowledges financial support from the CAS One Hundred Talent Fund and from NSFC grants 11673083 and 11333003.

Guoshoujing Telescope (the Large Sky Area Multi-Object Fiber Spectroscopic Telescope LAMOST) is a National Major Scientific Project built by the Chinese Academy of Sciences. Funding for the project has been provided by the National Development and Reform Commission. LAMOST is operated and managed by the National Astronomical Observatories, Chinese Academy of Sciences.

This work has made use of data from the European Space Agency (ESA) mission {\it Gaia} (\url{https://www.cosmos.esa.int/gaia}), processed by the {\it Gaia} Data Processing and Analysis Consortium (DPAC, \url{https://www.cosmos.esa.int/web/gaia/dpac/consortium}). Funding for the DPAC has been provided by national institutions, in particular the institutions participating in the {\it Gaia} Multilateral Agreement.

\end{document}